\documentclass[12pt]{article}

\usepackage[english]{babel}
\usepackage[dvips]{graphicx}

\title{Precision measurements, extra quark-lepton generations and 50 GeV
 neutrinos}
\author{V.A.Novikov \\ ITEP, Moscow, Russia \\}
\date{}

\def\ga{\mathrel{\mathpalette\fun >}}
\def\fun#1#2{\lower3.6pt\vbox{\baselineskip0pt\lineskip.9pt
\ialign{$\mathsurround=0pt#1\hfil##\hfil$\crcr#2\crcr\sim\crcr}}}

\begin{document}
\maketitle

\begin{abstract}
The existence of extra chiral generations with all fermions heavier 
than $m_Z$ is strongly disfavored by the precision electroweak data.
 However
the data still allow a few extra generations if the neutral leptons
have masses close to 50 GeV. Such heavy neutrino can be searched in the
reaction  $e^+ e^- \to N\bar{N}\gamma$. Existence of 50 GeV neutrinos makes
standard model Higgs boson invisible. \end{abstract}

\section{Introduction}\label{sec:intr}
The Standard Model (SM) constitutes three generations of quarks and leptons.
Nobody knows any deep reason for the number of
generation to be equal three. There are no theoretical arguments against
addition of extra sequential generations into SM.  So it would be
interesting to understand whether  such generations are allowed by
existing experimental data.   

 The direct experimental searches have found no trace of any 
extra quarks or leptons. The lower bounds on the masses  
of unobserved  generations are collected in the PDG tables \cite{1}. The
value of the  bounds are of the order of 100 GeV - they are fixed 
by the energy of the acting accelerators.    

There is also indirect way to study new generations with arbitrary heavy 
masses. To do that one has to study radiative corrections
and compare theoretical predictions with precision experimental data \cite{2} -
 \cite{7}. Indeed new particles are produced as
the virtual states in the propagation of the gauge bosons. These new loop
corrections into self-energies  of propagators ("oblique"
corrections) slightly change the predictions of the SM for low-energy physical
observables.  If the accuracy of the experimental data is better than the
value of these new corrections
one can exclude or discover new generations comparing theoretical predictions
with precision measurements.   
 
Consider this possibility in more details.
All set
of nearly 20 observables\footnote{ 
Decay  $Z \to b\bar{b}$
needs special consideration.}    that have been measured at LEP I, SLAC, Tevatron
and in $\nu N$ experiments (including the axial coupling $g_A$, the ratio
$R=g_V/g_A$, and the ratio $m_W/m_Z$) in the framework of the SM can be
determined in terms of three functions $V_i $(i=A,R,m) \cite{8}: 
\begin{equation}  m_W/m_Z = c +  \frac{3c}{32\pi s^2(c^2 - s^2)}
\bar{\alpha}V_m\;\; .  \label{46}  \end{equation}
 \begin{equation}
 g_{Al} = -\frac{1}{2} - \frac{3}{64\pi s^2c^2} \bar{\alpha} V_A\;\; ,
 \label{47}
 \end{equation}
 \begin{equation}
R \equiv g_{Vl}/g_{Al} = 1 - 4s^2 + \frac{3}{4\pi(c^2 - s^2)}
\bar{\alpha} V_R\;\; .
\label{48}
\end{equation}

The contribution of the new generations modifies the SM value of the
function $ V_i$ by  $\delta V_i$. 
It is important that interaction of gauge bosons
with new particles is universal, i.e. the gauge  coupling constants are the
same for any generations. So the loop corrections to the gauge
boson propagators can't be made arbitrary small, they are unambiguously fixed
by gauge couplings and by the masses of virtual particles. 
The second important
observation is that the matter in the SM is the chiral one, i.e. 
left-handed leptons and quarks belong to $SU(2)$ doublets
while their right-handed  companions  are
$SU(2)$ singlets. In this case heavy flavors do not decouple from the
low-energy physics even for very large masses, i.e. radiative corrections
remain finite when the masses of virtual particles go to infinity. 

For example in  the case 
of one extra generation $(UD)_L$, $(NE)_L$, $U_R$, $D_R$, $N_R$, $E_R$
with
fully degenerate doublets ($m_U=m_D=m_N=m_E$) and  ($m_U \gg m_Z$)
 one obtains "finite" corrections:
\begin{equation} \delta V_A = 0 \; , \;\; \delta V_R =-\frac{8}{9} \; , \;\;
\delta V_m =-\frac{16}{9}s^2 \;\; . \label{10}
\end{equation}

In the opposite case when $SU(2)$ is strongly violated  ($m_U \gg m_D$
or $m_U \ll m_D$)     all
corrections $\delta V_i$ are enhanced by the ratio of mass:
\begin{equation}
\delta V_i =\frac{|m_U^2 -m_D^2|}{m_Z^2} + \frac{1}{3}\frac{|m_N^2
-m_E^2|}{m_Z^2} \;\; .
\label{11}
\end{equation}

Thus extra heavy generations can be excluded if these additional 
contributions $\delta V_i$ exceed the discrepancy between the SM fit and
precision experimental data or can be "discovered" if they improve the fit.

In this way it was found \cite{1} -  \cite{6} that the existence of extra
chiral generations with all  fermions heavier than $M_Z$
is strongly disfavored by the precision
electroweak data (see section 2). The message of this talk is that there is a
small region in the parameter space where new generations could still exist.
Namely electroweak data are fitted nicely even by a few extra
generations, if one allows neutral leptons to have masses close to $50$ GeV
\cite{7} (see section 3).  Such neutrinos are still allowed by existing
experimental data. In section 4 we will consider the ways to detect heavy
$50$ GeV neutrinos. 

 During the discussions at this conference it was noted  by Valery Khoze
 \cite{VKh} that if heavy neutrinos do exist
the decay rate of  the Standard Model Higgs boson into the pair of heavy
neutrino would be about two order of magnitude higher than into $ b\bar{b}$,
i.e. the predominant decay mode of Higgs boson would be invisible.

\section{Heavy
generations and LEPTOP fit to experimental data}\label{sec:LEPTOP}

We compare theoretical predictions for extra generations with
experimental data \cite{EWWG}
with the help of the code LEPTOP \cite{10}.
In what follows we will assume that the mixing of the new generations with
the three old ones is small. Thus new fermions contribute only into
oblique corrections (vector boson self energies).

Heavy neutrinos need special consideration. There are two ways to make neutrino
massive. One can introduce right-handed neutrals $N_R$  and supply new
``neutrinos" with Dirac masses analogously to the case of charged leptons and
quarks. Another way is to construct  Majorana heavy neutrino. In
this paper we do not consider that possibility, our neutral lepton is a heavy
Dirac particle.

\begin{table}[h]
\caption{LEPTOP fit of the precision observables.}
\centering\label{tab:observables}
\begin{tabular}{|l|l|l|r|}
\hline 

Observ. & Exper.  & LEPTOP & Pull \\
           &  data   & fit    &      \\
\hline
$\Gamma_Z$ {\tiny[GeV]} &    2.4952(23) &  2.4964(16)  & -0.5   \\
$\sigma_h$ [nb] &   41.541(37)& 41.479(15)  & 1.7    \\
$R_l$ &   20.767(25)& 20.739(18)  & 1.1   \\
$A_{FB}^l$ & 0.0171(10)  &  0.0164(3)  & 0.7   \\
$A_{\tau}$ & 0.1439(42) &  0.1480(13)  & -1.0   \\
$A_e$ &  0.1498(48) &  0.1480(13) & 0.4  \\
$R_b$ &    0.2165(7) &   0.2157(1)  & 1.2   \\
$R_c$    &    0.1709(34)&   0.1723(1)  & -0.4   \\
$A_{FB}^b$  &    0.0990(20)&   0.1038(9)  & -2.4   \\
$A_{FB}^c$  &    0.0689(35)&   0.0742(7)  & -1.5   \\
$s_l^2$ {\tiny $(Q_{FB})$}   &    0.2321(10)  &   0.2314(2)  & 0.7   \\
\hline
$s_l^2$ {\tiny ($A_{LR}$)} & {\it 0.2310(3)}&  {\it  0.2314(2)}  & -1.5   \\
$A_b$ &  0.911(25) &  0.9349(1)  & -1.0   \\
$A_c$ &  0.630(26)&   0.6683(6)  & -1.5   \\
\hline
$m_W$ {\tiny [GeV]} & 80.434(37) & 80.397(23)  & 1.0   \\
$s_W^2$ {\tiny  ($\nu N$)} & 0.2255(21)& 0.2231(2) & 1.1 \\
\hline
$m_t$ {\tiny [GeV]}    & 174.3(5.1) &   { 174.0(4.2)} & 0.1\\
$m_H$ {\tiny [GeV]}    &   &  { $55^{+45}_{-26} $} &  \\
$\hat{\alpha}_s$ &           &  0.1183(27) &  \\
$\bar{\alpha}^{-1}$ & 128.88(9)           & 128.85(9) & 0.3  \\
{\small $\chi^2/n_{dof}$} & & 21.4/14 & \\
\hline
\end{tabular}
\end{table}
  We perform the four parameter ($m_t, m_H, \alpha_s, \bar{\alpha} $) fit to
18 experimental observables. The fitted parameters
together with the values of the predicted observables and their pulls
from the experimental data are given in the Table~\ref{tab:observables}.
The conclusion of the fit is quite evident: 
{\bf Standard Model fit of experimental data statistically is very good}.

There is only one cloud on the blue sky - the experimental value of the
forward-backward asymmetry in the  Z decay into the pair of b-quarks
$A_{FB}^b$ shows a hint for disagreement with Standard Model.
 \footnote{  With new BES and Novosibirsk data the accuracy of
$\bar{\alpha}^{-1}$ has to be improved in the nearest future. For
 value $\bar{\alpha}^{-1}=128.945(60)$
from ref.\cite{bolek} LEPTOP fit gives slightly   higher prediction for
the higgs mass   $m_H =     78^{+53}_{-32}$ GeV,
  $m_t =     174.1(4.5)$ GeV,
  $\alpha_s =     0.1182(27)$,
  $\bar{\alpha}^{-1}=128.927(58)$ and 
  $\chi^2 / ndf =21.1/14 $.}

Now we compare theoretical predictions for the case of
the presence of extra generations. The procedure is the following: 

First
we take
$m_D = 130$ GeV -- the lowest value allowed for the new quark mass from
Tevatron search \cite{11}  and take $m_U \ga m_D$.
As for the leptons from the extra generations, their masses are independent
parameters. To simplify the analyzes we start with
 $m_N = m_U$, $m_E = m_D$.
Any value of higgs mass above $113 GeV $
is allowed
 in our fits, however
$\chi^2$ appears to be minimal for $m_H=113$ GeV.

\begin{figure}[h]
\includegraphics
[width=0.5\textwidth,height=0.5\textheight]
{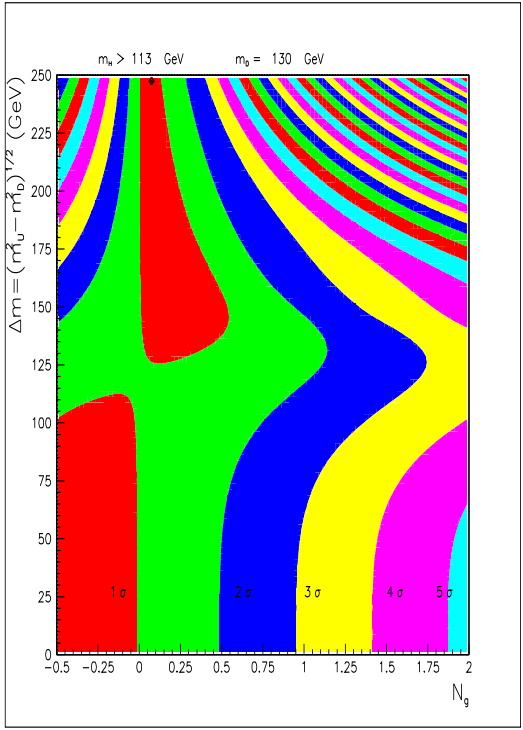}
\caption{
 Constraints on the number of extra generations $N_g$ and the mass
   difference in the extra generations $\Delta m$.
   The lowest allowed value  $m_D=130$ GeV from Tevatron search
   was used and $m_E=m_D$, $m_N=m_U$ was assumed.}
\label{fig:1}
\end{figure}

  In Figure  \ref{fig:1} the excluded
domains in coordinates ($N_g$, $\Delta m$) are shown 
(here $\Delta m = (m_U^2 - m_D^2)^{1/2}$ and the number of extra
generations $N_g$ is considered as a  continuous parameter).  Minimum   of
$\chi^2$ corresponds to $N_g =0.1$.  We see that one extra generation
corresponds to $2\sigma$ approximately.

We checked that similar bounds are valid for the general choice of heavy
masses of leptons and quarks. In particular we found that for 
 $m_N = m_D = 130$ GeV and $m_E = m_U$ one extra generation is excluded
at 1.5 $\sigma$ level, while for  $m_E = m_U = 130$ GeV and $m_N = m_D$
the limits are even stronger than in Fig.\ref{fig:1}.

The conclusion of this section is the following:

{\bf  Extra generations are excluded by 
the electroweak precision data, if all
extra fermions are heavy: $m \ga m_Z$}.

\section{"Light" heavy neutrino}\label{sec:heavy}

 Lower bounds on charged flavors $m_E$, $m_U$, $m_D$ are
approximately $100$ GeV-$130$ GeV.  However neutral lepton $N$
can be considerably lighter.

From LEP II searches of the decays $N\to
lW^{\ast}$, where $W^{\ast}$ is virtual while $l$ is $e$, $\mu$ or $\tau$, it
follows that $m_N > 70-80$ GeV for the mixing angle of the 4th generation
with three known generations larger than $10^{-6}$ \cite{12}.  The
quasi-stable neutral lepton $N$ with mixing angle less than $10^{-6}$ is 
allowed even for $m_N < 70$ GeV . In this case   neutrino $N$ escapes
detection from this LEPII search. The direct DELPHI  bound from the measurement
of the $Z$-boson width is much weaker: $m_N > 45$ GeV \cite{13}

We consider the case when new heavy neutrino has Dirac mass in the region
of $m_Z/2$.  For particles with masses of the order of $m_Z/2$ oblique
corrections drastically differ from what we have for masses $\ga m_Z$. 
In this case the Z boson state and $N \bar{N}$ state are practically degenerate
and even small mixing term can produce large change for eigenfunctions. In
other words renormalization of $Z$-boson wave function due to  $N \bar{N}$
intermediate state has to be large. 
 
It happens that  large wave function renormalization due to
"light" neutrino $N$  and contributions from "genuine heavy" $U,D$ and $E$
to $V_i$ have opposite sign. For very narrow region of the mass we have rather
good compensation .  Thus corrections to SM 
formulas $\delta V_i$  are small and we have the kind of heavy generation
conspiracy.

\begin{figure}[h]
\includegraphics
[width=0.65\textwidth,height=0.5\textheight]
{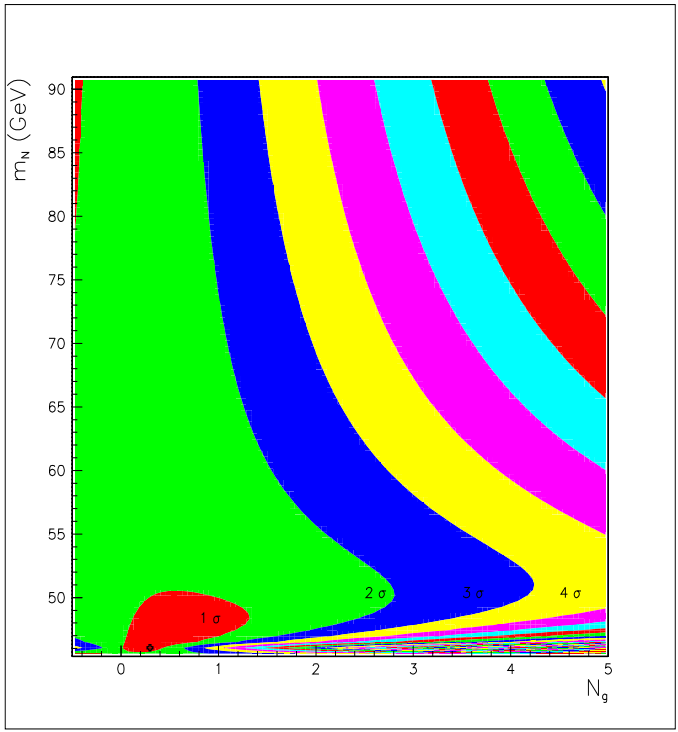}
\caption{Constraints on the number of extra generations $N_g$ and the mass
   of the neutral heavy lepton $m_N$.
   The values  $m_U=220$ GeV, $m_D=200$ GeV, $M_E=100$ GeV
   were used}

\label{fig:2}
\end{figure}

  As an example we take $m_U =
220$ GeV, $m_D = 200$ GeV, $m_E = 100$ GeV and draw exclusion plot in
coordinates ($m_N, N_g$), see Fig.\ref{fig:2}.
  From this  plot it is clear that for the case of fourth
generation with $m_N \approx 50$ GeV description of the data is not worse
than for the Standard Model and that even two new generations with $m_{N_1}
\approx m_{N_2} \approx 50$ GeV are allowed within $1.5 \sigma$.

 The conclusion of this section is the following:

{\bf  If the neutral lepton $N$ has mass around 50 GeV the new generations are
not excluded by  the electroweak precision data}.

\section{The direct search of the heavy neutrino}\label{sec:search}

The direct search of the heavy neutrino is possible in 
 $e^+e^-$-annihilation.
As was proposed a long time ago \cite{14}, the cross section of
$e^+e^-$-annihilation into an invisible final state can be inferred from
observation of initial state bremsstrahlung, i.e from
$e^+e^-$-annihilation into a pair of heavy neutrinos with the emission
of initial state bremsstrahlung photon
\begin{equation}
e^+ e^- \to \gamma + \; \mbox{\rm N} \bar{\mbox{\rm N}}
\label{eq:1}
\end{equation}
The main background is the production of the pairs of conventional
neutrinos with initial state bremsstrahlung photon
\begin{equation}
e^+ e^- \to \gamma + \; \nu_i \bar\nu_i
\label{eq:2}
\end{equation}
where $i = e, \mu, \tau$. These background neutrinos are produced
in decays of real and virtual $Z$. In case of  $\nu_e
\bar\nu_e$, two mechanisms contribute, through $s$-channel $Z$ boson
and from $t$-channel exchange of $W$ boson.
We calculated the signal and background distributions
and rates \cite{mn50} using CompHEP \cite{comphep} computer code.

\begin{figure}
\includegraphics
[width=0.4\textwidth,height=0.4\textheight]
{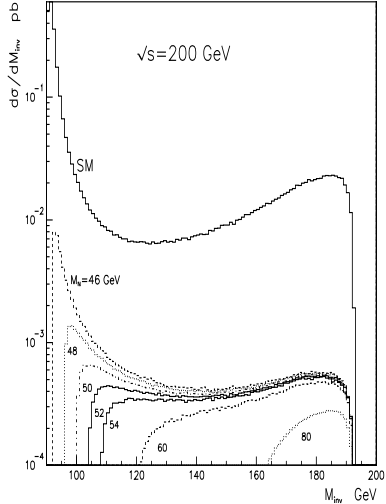}
\caption{ $d\sigma/d M_{inv}$ (in pb) for 
Standard Model and for the different values of $m_N$.}
\label{fig:3}
\end{figure}

In Fig.\ref{fig:3} the distribution on
``invisible" mass $M_{inv}$ (invariant mass of the neutrino 
pair)is represented for SM background and the 
$N\bar N$ signal for $\sqrt{s}=200$
GeV and different values of $N$ masses, $M_N=46-100$ GeV. Here we applied
kinematical cuts on the photon polar angle and transverse momentum,
$|\cos\vartheta_\gamma|<0.95$ and $p_T^\gamma>0.0375\sqrt{s}$, being the
ALEPH selection criteria \cite{ALEPH}.
The photon detection efficiency 74\% is assumed.
For highest significance of the $N\bar N$ signal, evaluated as
$N_S/\sqrt{N_B}$, one should include whole interval on $M_{inv}$ allowed
kinematically, so we applied  $M_{inv}>2m_N$ cut.

\begin{figure}
\includegraphics
[width=0.4\textwidth,height=0.4\textheight]
{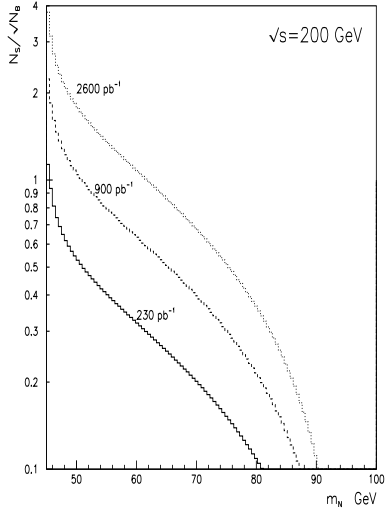}
\caption{
$N\bar N$ signal significances at LEP-2 at different statistics
as function of the neutrino mass.}
\label{fig:s-mN}
\end{figure}

On Fig.\ref{fig:s-mN} the signal
significances
are represented as a function
of $m_N$. One can derive that only the analysis based on combined data from
all four experiments both from 1997-1999 runs ($\sqrt{s}=182-202$ GeV) and
from the final run, in total $\sim 2600$ pb$^{-1}$, can exclude at 95\%
CL the interval of $N$ mass up to $\sim 50$ GeV.

Another possibility is to search for 50 GeV neutrino at the future
TESLA $e^+ - e^-$ electron-positron linear collider.
The increase in energy
leads to the decrease both of the signal and the background,
but it is 
compensated by the proposed increase of
luminosity
of 300 ${\rm fb}^{-1}$/year \cite{TESLA}.
Further advantage of the linear collider is the possibility to use
polarized beams.
 This is important in suppressing the cross section of $e^+ e^-
\to \nu_e\bar{\nu}_e\gamma$ as this reaction goes mainly through the
$t$-channel exchange of the $W$ boson.
 However, even without exploiting the beam
polarization the advantage of TESLA in the total number of events is
extremely important. Thus, Standard Model is expected to give approximately
0.3 million single photon events for $M_{inv}>100$ GeV while the number of
50 GeV neutrino pairs would be about 4000. 

Although the signal over background ratio is still small
(2.3-0.5\% for $m_N=45-100$ GeV correspondingly) the significance of the
signal is excellent, higher than 5 standard deviations for $m_N<60$ GeV.
 
The conclusions of this section are the following:

  Combined data from four LEP II experiments can exclude at $95\%$ CL 
neutrino with  $M_N < 50$.

   Future collider TESLA in one year run can exclude at $95\%$ CL the whole
region of $m_N$ or can discover neutrino
within 5 standard deviations
with $m_N<60$ GeV. 

\section{ Conclusions}

 The existence of extra chiral generations with all fermions heavier 
than $m_Z$ is disfavored by the precision electroweak data.
 A few extra generations with "light" neutral leptons, i.e.
with mass $M_N$ close to 50 GeV,  are not excluded yet by existing data.
Such heavy neutrino can be searched in the reaction  $e^+ e^- \to
N\bar{N}\gamma$.

 After this talk Valera Khoze made simple and important 
observation, that 
$50$ GeV neutrino makes Standard Model Higgs boson invisible, that require
a special strategy for its search  \cite{VKh}.

\section{Acknowlegements}

I am grateful to the organizers of  La Thuile conference, particularly
to Mario Greco, for their outstanding hospitality and for a very pleasurable
and thought provoking conference.

\end{document}